\documentclass{aastex}
\usepackage{spr-astr-addons}
\usepackage{url}

\RequirePackage{color}

\begin{document}

\title{The 5th edition of the {\it Roma-BZCAT}. A short presentation}
\slugcomment{}
\shorttitle{The 5th edition of the Roma-BZCAT}
\shortauthors{E. Massaro et al.}

\author{E. Massaro\altaffilmark{1,2}} 
\altaffiltext{1}{INAF, Istituto di Astrofisica e Planetologia Spaziali, Roma, Italy}
\altaffiltext{2}{In Unam Sapientiam, Roma, Italy} 
\and 
\author{A. Maselli\altaffilmark{3}} 
\altaffiltext{3}{INAF, Istituto di Astrofisica Spaziale e Fisica Cosmica, Palermo, Italy} 
\and
\author{C. Leto\altaffilmark{4}}
\altaffiltext{4}{ASI Science Data Center, Roma, Italy} 
\and
\author{P. Marchegiani\altaffilmark{5}} 
\altaffiltext{5}{School of Physics, University of the Witwatersrand, Johannesburg Wits, 2050, South Africa} 
\and
\author{M. Perri\altaffilmark{4,6}}
\altaffiltext{4}{ASI Science Data Center, Roma, Italy} 
\altaffiltext{6}{Osservatorio Astronomico di Roma, INAF, Monteporzio Catone, Italy} 
\and
\author{P. Giommi\altaffilmark{4}} 
\altaffiltext{4}{ASI Science Data Center, Roma, Italy} 
\and
\author{S. Piranomonte\altaffilmark{6}} 
\altaffiltext{6}{Osservatorio Astronomico di Roma, INAF, Monteporzio Catone, Italy}

\begin{abstract}
The 5th edition of the {\it Roma-BZCAT} Multifrequency Catalogue of Blazars is available in a
printed version and online at the ASDC website ({\sf http://www.asdc.asi.it/bzcat}); it is also in the 
~NED \\ database. 
It presents several relevant changes with respect to the past editions which are briefly described
in this paper.
\end{abstract}


\section{Introduction}

The work on the {\it Roma-BZCAT} (Massaro et al. 2009), which is a list of carefully 
checked blazars, started more than 10 years ago. 
It was originally conceived for developing a complete database from catalogue and literature
data for the identification of counterparts to high energy sources.
The Fermi-LAT collaboration, in fact, used it at this purpose in various $\gamma$-ray 
source catalogues, like 1FGL (Abdo et al. 2010a), 1LAC (Fermi LAT AGN Catalog, Abdo et al. 
2010b) and the subsequent 2FGL (Nolan et al. 2012) and 2LAC (Ackermann et al. 2011).
However, considering that the sources reported in the catalogue were divided into a few classes
having a remarkable homogeneity, it was also successfully applied in new blazar researches 
and provided powerful selection criteria for surveys in other electromagnetic bands.

The {\it Roma-BZCAT} is now at the 5th Edition (Massaro et al. 2014) which contains 
coordinates and multifrequency data of 3561 sources, about 30\% more than in the 1st
edition, either confirmed blazars or exhibiting characteristics close to this type of 
sources.
With respect to the previous editions, this new edition has relevant changes in the 
sources' classification and has a new format for the notes in the tables.
We emphasize that all the sources in the Roma-BZCAT have a detection
in the radio band. 
Moreover, a complete spectroscopic information is published and could
be accessed by us for all of them, with the exception of BL Lac
candidates.
Consequently, peculiar sources as the so called ``radio quiet BL Lacs'', which are reported 
in some other catalogues, are not included here because of possible contamination with hot stars
and other extragalactic objects.

In this paper we summarize the major changes with respect to the previous editions and give 
some indications on the use of the online version.
For a more complete description we address to the catalogue printed version (Massaro et al. 2014).

\begin{figure*}[ht]
\begin{centering}
\includegraphics[height=8.0cm]{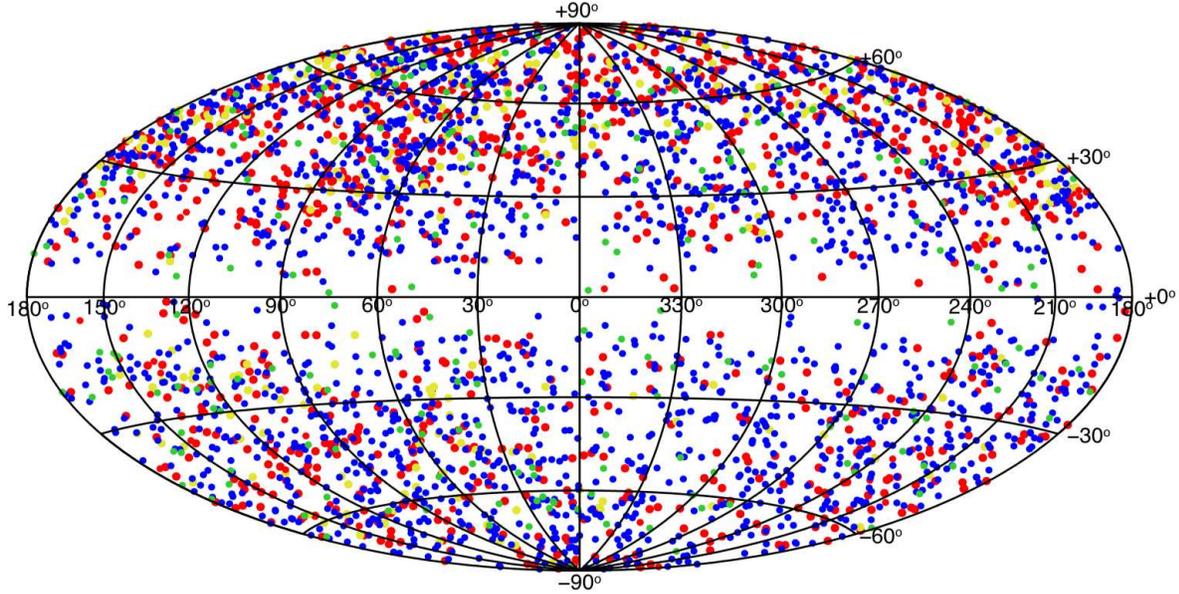}
\caption{%
Hammer-Aitoff projection in galactic coordinates of the sky distribution of 
blazars in the 5th Edition of the {\it Roma-BZCAT}. 
The dots' colours correspond to the following sources' types: BL Lacs and candidates (red), 
BZGs (yellow), FSRQs (blue),  blazars of uncertain classification (green) [from Massaro et al. 2014].
} 
\end{centering}
\end{figure*}

\section{Types of blazars and source naming}

Traditionally, blazars come in two main flavors: BL Lac objects and Flat 
Spectrum Radio Quasars (FSRQ), depending on the width and strength of the 
emission lines in their optical spectrum.
Variability and other uncertainties, however, raised the problem of an accurate 
blazar definition, an issue that is still open. 
Since the first edition, therefore, we divided blazars into these two main classes,
originally named as BZB and BZQ, respectively.
It was clear, however, that there is a fraction of Active Galactic
Nuclei (AGN) exhibiting blazar properties mixed to other features
which make difficult the classification in the two previous types.
Thus these sources were included in the {\it Roma-BZCAT} as blazars of uncertain type 
or BZU, together with sources with data too poor for a safe classification.
New studies of the blazar population properties, also based on the results of high 
energy observations (Massaro, Nesci and Piranomonte 2012), suggested us the definition
of the new class of BZG blazars (or blazar candidates), whose multifrequency emission 
exhibits some properties of blazars but appears dominated by the host galaxy contribution 
in the optical-UV.
It is not clear if BZG objects, which are characterized by redshift values generally 
lower than 0.3, are all genuine blazars, for instance in a rather
quiescent state considering the high variability that they can present, or are moderately
bright AGNs whose non-thermal emission does not present evidence 
for relativistic beaming, or again misclassified sources.
Only very accurate observational investigations of these sources can provide more 
stringent criteria for a better classification. 

In the 5th Edition we use similar denomination of blazars adopted in the previous editions. 
Each blazar is identified by a code, with 5BZ for all blazars, a fourth letter that
specifies the type, followed by the truncated equatorial coordinates (J2000).
We introduced the edition number before the letters BZ to avoid possible confusion due to 
the fact that several sources changed their old names because of the new adopted classification.

The codes are:
\begin{itemize}
\item
 {\bf 5BZB}: BL Lac objects, used for AGNs with a featureless optical spectrum, or
having only absorption lines of galaxian origin and weak and narrow emission
lines;
\item
 {\bf 5BZG}: sources, usually reported as BL Lac objects in the literature,
but having a spectral energy distribution (SED) with a significant dominance of the galaxian 
emission over the nuclear one;
\item
 {\bf 5BZQ}: Flat Spectrum Radio Quasars, with an optical spectrum showing broad
emission lines and dominant blazar characteristics;
\item
 {\bf 5BZU}: blazars of Uncertain type, adopted for a small number of sources having
peculiar characteristics but also exhibiting blazar activity: for instance, occasional
presence/absence of broad spectral lines or other features, transition objects between
a radio galaxy and a BL Lac, galaxies hosting a low luminosity blazar Nucleus, etc.
\end{itemize}

The 5th edition contains 1151 BZB sources, 92 of which are reported as candidates
because we could not find their optical spectra in the literature, 1909 BZQ sources,
274 BZG sources and 227 BZU objects.

\section{Multifrequency data}
For each source, in addition to the J2000 coordinates mostly derived from VLBI (Titov \& Malkin 2009; Titov et al. 2011; Petrov \& Taylor 2011), WISE and optical
databases like SDSS, the following data are also given:

\noindent
--~the apparent magnitude $R$ from USNO B1 or $r$ from SDSS DR10, or in other bandpasses when these data
are not available;

\noindent
--~radio flux density from NVSS (Condon et al. 1998) or FIRST (White et al. 1997) at 1.4 GHz or at 0.843 GHz from SUMSS (Mauch et al. 2003) when the former 
ones are not available; 
         
\noindent
--~radio flux density at 4.85 GHz from GB6 (Gregory et al. 1996) or PMN (Wright et al. 1994), for a few other sources other bands are 
used as specified by notes;

\noindent
--~microwave flux density, mostly at 143 GHz, from PLANCK (Planck Compact Source Catalogue Public Release 1, Planck Collaboration, 2013) catalogues;

\noindent
--~the 0.1--2.4 keV X-ray flux from ROSAT archive or Swift-XRT catalogues, in other cases as specified by notes;

\noindent
--~hard X-ray flux (15--150 keV) from Palermo BAT Catalogue (Cusumano et al. 2010), in other cases as specified by notes; 

\noindent
--~$\gamma$-ray flux from 1FGL or 2FGL catalogues;
 
\noindent
--~the redshift.

When data for the considered band were not found in catalogues or in the literature a sign - is
reported, while a sign + indicates that some data are available but an additional analysis is necessary
to be inserted in the catalogue. 
In the latter case the corresponding note gives information on the instrument archive where data 
can be recovered. 

\section{On-line version and scientific tools}
The on-line version of the {\it Roma-BZCAT} provides access to useful tools developed at
ASDC that can be easily accessed by clicking on the Data Explorer button. 
For instance it is possible to build sky maps of catalogued sources in the region surrounding
the selected blazar or retrieve optical and radio images at different size scale. 
A large series of catalogues in many electromagnetic bands is available and all major databases
can be accessed in a transparent way, including bibliographical services.
A useful tool is the ASDC SED builder: the user can obtain time resolved SEDs of
a selected source (see Fig. 2) from a collection of available data and
possibly add his own dataset. Then, he can also calculate some
emission models by means of a synchrotron self-Compton code. Finally,
the user can evaluate the spectral parameters of the sources for
investigating the expected fluxes in other bands according to
different criteria.

\begin{figure}[t]
\includegraphics[width=8.4cm]{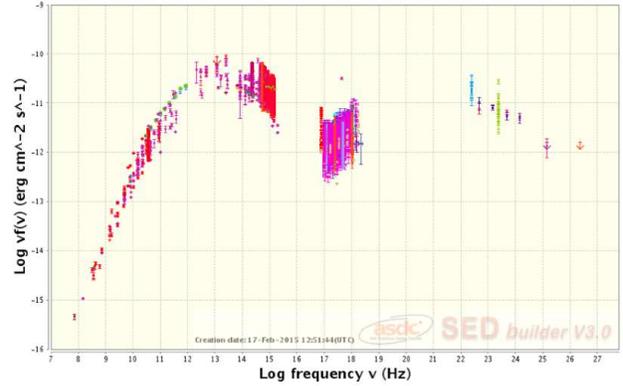}
\caption{%
The spectral energy distribution of the very well studied BL Lac object OJ 287/5BZB J0854+2006
obtained by means of the ASDC tool.
Used symbols are described in the on-line page.}
\end{figure}

\section{Conclusion}

The {\it Roma-BZCAT} is likely the most complete list of blazars, useful for identifying the 
counterparts to high-energy sources and for investigating the populations of extragalactic sources
in other frequency bands such as in the microwaves or in the far and mid infrared.
Moreover, it provides a large database of SEDs for different types of blazars useful for studying 
radiation mechanisms and relativistic beaming effects.

The completeness of the catalogue is an open issue, because the sky coverage of the used
surveys is not uniform. 
In the past editions we noticed that the number of blazars in the Northern sky was higher than
in the Southern part.
This situation is partially changed with the new large surveys, like AT20G (Massardi et al. 2011), 
as shown by the declination distributions of BL Lac objects and FSRQs reported in Fig. 3.
The N-S asymmetry is well apparent for the former sources while the latter ones does not show any
significant effect.
This difference is due to the fact that many BL Lac objects classified as HBL (High frequency peaked
BL Lac, see Padovani and Giommi 1995) have radio flux densities at 1.4 GHz lower than about 50 mJy
and are not detected in the surveys of the Southern sky, while only a very small fraction of FSRQs have
flux densities lower than this value.
Moreover, one has to consider that a relevant number of new discovered BL Lacs was identified using 
spectroscopic observations available in the SDSS, and that a similar facility is not present in
the southern sky observatories.
We expect that the combination of radio and optical data with other multifrequency surveys, as for example with 
the mid-infrared colours from the AllWISE catalogue (Cutri et al. 2013), as recently proposed by Massaro 
et al. (2011, 2013) and D'Abrusco et al. (2012, 2013), will increase the efficiency for discovering 
new HBL objects.

Users who will make use of the {\it Roma-BZCAT} (5th Edition) in a
publication, are kindly requested to acknowledge the source of the
information by referencing the present paper or the book which gives a
full description of the catalogue.

\begin{figure}[t]
\includegraphics[width=\columnwidth]{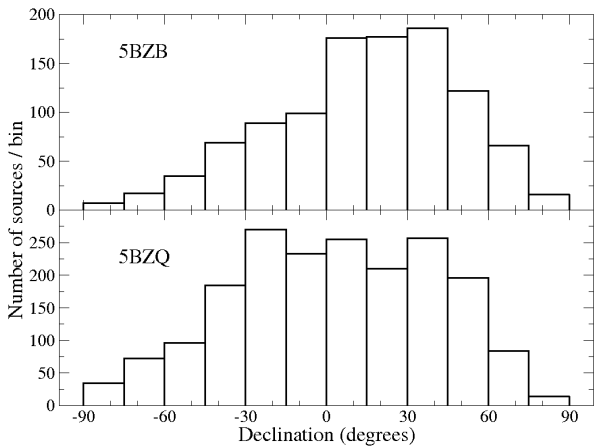}
\caption{%
The Declination distributions of BL Lac objects (upper panel) and of Flat Spectrum Radio Quasars
(lower panel) in the {\it Roma-BZCAT}. 
} 
\end{figure}

\acknowledgments
This work started when one of the authors (E.M.) was at the Universit\`a di Roma La Sapienza. 
We made large use of the Sloan Digital Sky Survey (SDSS), the NED database and other 
astronomical catalogues distributed in digital form (Vizier and Simbad)
at Centre de Dates astronomiques de Strasbourg (CDS) at the Louis Pasteur University.
This work has made use of the Palermo BAT Catalogue and database operated at INAF-IASF Palermo. 
Spectra from the Final Release of 6dFGS have been used in our blazar diagnostics.

The use of the online version of the {\it Roma-BZCAT} and the scientific tools developed 
at the ASI Science Data Center (with the supervision of Paolo Giommi) were fundamental
for the catalogue revision. 
We are therefore grateful to the ASDC technical staff for the excellent work carried out 
to support the on-line version of the {\it Roma-BZCAT}.

P.M. acknowledges support from the DST/NRF SKA post-graduate bursary initiative.
This work has been partially supported by research funds of Universit\`a di Roma La Sapienza 
and by the ASI grant to INAF to support the scientific activities of the ASDC.

\end{document}